\documentclass[sigconf]{acmart}

\usepackage{booktabs} % For formal tables

\setcopyright{rightsretained}
\makeatletter
\renewcommand\@formatdoi[1]{\ignorespaces}
\makeatother

\acmISBN{}

\acmConference[]{Arxiv}{November}{27th} 
\begin{document}
\title[Concept-Based Hypertext for Case-Based Retrieval]{A Concept-Centered Hypertext Approach to Case-Based Retrieval}
%\titlenote{Produces the permission block, and
%  copyright information}
%\subtitlenote{The full version of the author's guide is available as
  %\texttt{acmart.pdf} document}

\author{Stefano Marchesin}
\affiliation{
  \institution{Department of Information Engineering, University of Padua }
  \streetaddress{Via Gradenigo 6/b}
  \city{Padua} 
  \state{Italy} 
  \postcode{35131}
}
\email{stefano.marchesin@unipd.it}

\renewcommand{\shortauthors}{S. Marchesin}

\thanks{\textbf{The paper is an extended version of the work presented at SIGIR 2018 \cite{marchesin-2018}.}}

\begin{abstract}
The goal of case-based retrieval is to assist physicians in the clinical decision making process, by finding relevant medical literature in large archives. We propose a research that aims at improving the effectiveness of case-based retrieval systems through the use of automatically created document-level semantic networks. The proposed research tackles different aspects of information systems and leverages the recent advancements in information extraction and relational learning to revisit and advance the core ideas of concept-centered hypertext models. We propose a two-step methodology that in the first step addresses the automatic creation of document-level semantic networks, then in the second step it designs methods that exploit such document representations to retrieve relevant cases from medical literature. For the automatic creation of documents' semantic networks, we design a combination of information extraction techniques and relational learning models. Mining concepts and relations from text, information extraction techniques represent the core of the document-level semantic networks' building process. On the other hand, relational learning models have the task of enriching the graph with additional connections that have not been detected by information extraction algorithms and strengthening the confidence score of extracted relations. For the retrieval of relevant medical literature, we investigate methods that are capable of comparing the documents' semantic networks in terms of structure and semantics. The automatic extraction of semantic relations from documents, and their centrality in the creation of the documents' semantic networks, represent our attempt to go one step further than previous graph-based approaches.
\end{abstract}




\maketitle

\section{Motivation for the proposed research}
\label{sec:motivation}
The volume of medical literature published every year keeps growing at a very fast pace. In terms of performance, this poses a real challenge to clinicians\footnote{Clinician is a general term that encompasses every medical position that involves patients; therefore, the responsibilities for clinicians vary depending on the job title. Clinicians can be medical assistants, physicians, surgeons, counselors, psychiatrists, and so on. In a nutshell, clinicians' responsibilities are diagnosing illnesses and administering treatment to patients through either medicine or procedures.} who need to examine such huge volume of literature. Indeed, large databases of published medical research can support clinicians in the clinical decision making process, by providing them with relevant information for a given case. However, the time required to retrieve relevant information from these databases using standard systems is often prohibitive, turning them into cumbersome resources for clinicians. Therefore, an increase in interest for systems that help clinicians to make clinical decisions has emerged. Such systems are known as Clinical Decision Support (CDS) systems.\par 

A CDS system is designed to assist clinicians in providing patient care by producing effective and timely knowledge that can help in the decision making process. One of the tasks of CDS systems is to retrieve, given a medical case of interest, highly related medical literature that could aid clinicians in formulating diagnoses or deciding treatments for the case at hand. This side of CDS systems is known as case-based retrieval. Therefore, the goal of case-based retrieval is to assist clinicians in the clinical decision making process by finding, in large collections, medical literature that most resembles a given case. Cases are described by reports, usually composed of a textual description on the medical findings and the related medical images. \par 

Case-based retrieval presents some unique challenges and peculiarities. Since queries are medical reports, this search task differs radically from other more common search domains, like Web search. In fact, clinical reports are substantially longer compared to Web search queries. Furthermore, having cases often a narrative structure, case-based retrieval differs also from those search domains where queries are long but keyword based, like legal or patent search. Last but not least, due to the limited time that clinicians can afford spending in reading literature while practicing, case-based retrieval strongly favors precision over recall \cite{burke_etal-2004}. \par 

Clinical reports are often created with minimal consideration for any subsequent computational analysis of the underlying concepts. Therefore, case-based retrieval systems have to process a variety of literature styles written with a wide domain-specific vocabulary, comprised of many synonyms and context specific expressions. This fact makes the comparison of clinical reports challenging, as reports lack a representation in a space where their similarity can be measured quantitatively. \par

Ontologies and thesauri have been often exploited by retrieval systems as a mean to overcome such issues --- concept-based Information Retrieval (IR) is the most notable example of this. However, the semantic relations within these knowledge sources have been only partially leveraged, not fully exploiting their power to semantically represent relations between documents' concepts. The semantic relations expressed within documents carry high informative power that can be the turning point to create more semantic-aware concept-based models. \par 

The combination of recent developments in Information Extraction (IE) and the availability of unparalleled medical resources, thus offers us the opportunity to develop new techniques that better capture the semantics within medical reports. Along with IE techniques, relational learning models can also be employed. Relational learning studies methods for the statistical analysis of graph-structured data. The main objectives of relational learning include prediction of missing edges, prediction of properties of nodes, and clustering nodes based on their connectivity patterns. \par 

Bearing in mind the centrality of providing a higher semantic understanding of the clinical reports' contents, the remainder of the paper is organized as follows: Section \ref{sec:related_work} presents the background and related work, Section \ref{sec:research_description} describes the purposes and aims of the proposed research, Section \ref{sec:research_methodology} discusses the proposed methodology and Section \ref{sec:research_issues} provides an outlook on future challenges. 

\section{Background and related work}
\label{sec:related_work}
The type of queries submitted to case-based retrieval systems --- and the need for results that are similar to the issued query --- relate case-based retrieval to several research directions in the literature, including: query-by-example (typically used in image retrieval \cite{ballerini_etal-2009}), item-to-item recommendation \cite{sarwar_etal-2001} and search-by-ideal-candidate \cite{ha-thuc_etal-2016} (used by LinkedIn in job search). \par 

Another related research direction is concept-based IR. Concept-based IR aims at making use of external knowledge sources (like thesauri and ontologies) to provide additional knowledge and context that may not be explicit in a document collection and users' queries. Concept-based IR represents both documents and queries using semantic concepts, instead of (or in addition to) keywords, and performs retrieval in that concept space. Concept-based methods can be categorized in two types: (i) methods that use concepts throughout the entire process, in both indexing and retrieval stages \cite{egozi_etal-2011}, and (ii) methods that apply concept analysis in one stage only, such as concept-based query expansion in \cite{grootjen_van-der-weide-2006} --- which is a simpler but less accurate solution. The approach we adopt can be considered as an extension of (i) to semantic relations between concepts.\par

In the biomedical domain --- where the authoritative and curated ontologies can provide a valuable source of knowledge --- concept-based approaches demonstrate consistent improvements over classic keyword-based systems. For instance, in \cite{koopman_etal-2012} queries and documents are transformed from their term-based originals into medical concepts as defined by the SNOMED CT ontology\footnote{\url{https://www.snomed.org/}}. In addition, parent-child 'is-a' relationships between concepts are used to weight documents that contain concepts subsumed by the query's concepts. In \cite{limsopatham_etal-2013}, the authors proposed a method to represent medical records and queries by focusing only on medical concepts essential for the information need of a medical search task. In \cite{limsopatham_etal-2013a}, queries are expanded by inferring additional conceptual relationships from domain-specific resources as well as by extracting informative concepts from the top-ranked medical records.\par 

The field of Biomedical Information Extraction (BioIE) has been active for many years and is highly relevant for clinical decision support. In \cite{liu_etal-2016} the authors focused on reviewing the recent advances in learning-based approaches for BioIE tasks. BioIE tasks comprise entity linking \cite{zheng_etal-2015}, event identification \cite{ananiadou_etal-2010} and relation extraction \cite{uzuner_etal-2011,wang_fan-2014}. The i2b2 tranSMART Foundation\footnote{\url{https://www.i2b2.org/index.html}} provided different challenges to evaluate BioIE systems on these tasks.\par 

Regarding relational learning, \cite{nickel_etal-2016} discusses two different kinds of relational models, both big data scale compliant. The former are based on latent feature models such as tensor factorization and multi-way neural networks \cite{nickel_etal-2011,bordes_etal-2013}. The latter are based on mining observable patterns in the graph \cite{lao_cohen-2010}. Combining these latent and observable models can improve the modeling power at decreased computational cost. In Google's knowledge vault project \cite{dong_etal-2014}, the combination of relational learning models and IE methods is performed for the construction of a knowledge graph from Web sources. \par

Graph-based models have been applied in IR since the early work of Minsky on semantic IR \cite{minsky-1969}, which was followed by several variants of conceptual IR and knowledge-based IR. With the advent of the Web, and in particular thanks to the seminal works of \cite{page_etal-1999,kleinberg-1999}, graph models flourished again. More recent works relate graph theoretic approaches to IR in the context of social or collaborative networks and recommender systems \cite{craswell_szummer-2007,konstas_etal-2009}. \par

The approaches described above usually build the graph out of the main components of an IR process (e.g. documents/queries/users). Instead, in \cite{blanco_lioma-2012} the authors build the graph out of the individual terms contained in a document. Hence, the object of their representation is not the IR process per se. The same approach has been followed by \cite{koopman_etal-2012a}, where Blanco's model \cite{blanco_lioma-2012} has been adapted to a concept representation of documents, in order to capture the dependencies between concepts found in medical free-text. Finally, in \cite{koopman_etal-2016} a graph inference retrieval model is presented  that integrates structured knowledge resources, statistical IR methods and inference in a unified framework.

\section{Description of proposed research}
\label{sec:research_description}
The proposed research has the objective of improving the effectiveness of case-based retrieval systems for clinical decision support. The research is twofold, comprising of a document graph creation phase and a case-based retrieval phase. Due to this dual nature, two main questions drive the research.

\begin{itemize}
\item[(i)] How can clinical cases and medical literature be represented in such a way that the semantic, and authoritative, information that lies within them can be connected and leveraged to the best?
\item[(ii)] How can be leveraged such semantic representations of clinical cases and medical literature in such a way that, given a query case, we can effectively return related clinical cases or medical literature?
\end{itemize}

(i) raises the question of how to represent documents using structures that allow the semantic and authoritative information to be explicitly brought out. Thus, our research presents, as a first step, the study and design of methods for extracting semantic information contained within documents. The principal constituents of the semantic information are authoritative concepts and relations between them. In order to extract authoritative concepts and semantic relations from unstructured medical reports it is necessary to leverage external knowledge sources. Therefore, the research is directed towards the use of these external sources, like knowledge bases, ontologies and thesauri. \par 

Once we have extracted concepts and relations, the second step is to provide a document representation that is well suited to express such semantic information contained within the document. In this way, the document representation can achieve a finer semantic level, by means of authoritative concepts and semantic relations describing the underlying document. Due to the nature of the information extracted by IE techniques, that is concepts and relations, the document representation is well suited to be conceived as a graph structure, precisely - a semantic network \cite{sowa-2014}. A semantic network is a graph structure for representing knowledge in patterns of interconnected nodes (concepts) and edges (relations). The main feature of semantic networks is the declarative graphic representation that can be used either to represent knowledge or to support automated systems for reasoning about knowledge. \par

The research proposed for (i) leads the research for (ii) to focus on leveraging the documents' semantic networks for case-based retrieval. 
Leveraging documents' semantic networks in both indexing and retrieval stages is the key element to bring case-based retrieval to a higher semantic level. Therefore, semantic similarity measures that are capable of comparing and relating such semantic networks one with each other will be studied and designed. The analysis of documents' semantic networks can be divided into two approaches: an explicit, or graph-based, approach and an implicit, or latent-based, approach. In short, the former makes use of the topological structure of the networks, along with semantic paths between the nodes (i.e. concepts), in order to define similarity measures that take into consideration the semantic interconnectivity between the elements of the network \cite{lao_cohen-2010, spanier_etal-2017}. The latter leverages graph kernels \cite{shervashidze_etal-2011, shervashidze_etal-2009} and embedding models \cite{narayanan_etal-2017, narayanan_etal-2016}, that can learn latent (low-dimensional) representations of graphs. Such latent representations can then be compared to find the highest correspondences between the query case and the medical literature. \par 
 
We believe that our research can improve case-based retrieval systems and also, as a side effect, tackle the semantic gap. In fact, the medical language amplifies some specific challenges of the semantic gap. Vocabulary mismatch is more accentuated and the interdependence between terms is greater than in other domains. Therefore, providing a finer semantic representation of medical literature, by building documents' semantic networks out of authoritative concepts and semantic relations, can reduce vocabulary mismatch and can better represent the interdependence between medical concepts within text. 

\section{Research methodology and proposed experiments}
\label{sec:research_methodology}
The starting point for the proposed research are the works on concept-centered hypertext models by Agosti \cite{agosti_crestani-1993,agosti_marchetti-1992,agosti_etal-1992}. The core ideas of these works are here revisited and leveraged to address the task of case-based retrieval. The description of the methodology is divided in three parts: a first document graph construction part, which describes the methodology for the automatic construction of documents' semantic networks; a second case-based retrieval part, which describes the methodology proposed to leverage such semantic representations for the retrieval of medical literature; and a third evaluation part, which describes the evaluation of the methods proposed in the first and second part.

\subsection{Document Graph Construction}
\label{subsec:document_graph_construction}
The research focused on a preliminary study of state-of-the-art methods and techniques related to entity linking, relation extraction and link prediction. An investigation of knowledge bases, ontologies and thesauri has also been performed. \par 

The method we propose for the automatic creation of documents' semantic networks is a combination of three different techniques: entity linking; relation extraction and link prediction. The first two techniques represent the core of the document graph creation, while link prediction comes into play in a second phase, where it is used to enrich the created graph with additional unknown relations. In fact, with link prediction, indirect linkages between source nodes and target nodes can be analyzed, and previously unknown relationships can be discovered. \par 

By extracting semantic relations from text, and by inferring additional connections only between extracted concepts --- as opposed to \cite{koopman_etal-2012a,koopman_etal-2016} --- the graph representation is more adherent to the contents that are explicitly stated in the document itself. \par 

For the link prediction part, we decided to exploit TransE \cite{bordes_etal-2013} as it is one of the most solid and easy-to-train algorithms for link prediction. In our work we are interested in predicting the most likely relations that exist between two concepts (nodes). Thus, the role of link prediction can be twofold. Its primary use is to enrich the document graph with previously unknown relationships between nodes (concepts), as already mentioned. However, as a secondary task, link prediction techniques can also be used to strengthen the confidence score of extracted relations, following the approach presented in \cite{dong_etal-2014}. \par 

For the entity linking part we tested different open-source tools. We decided to adopt MetaMap\footnote{\url{https://metamap.nlm.nih.gov/}}, the most authoritative tool to detect medical entity mentions in free-text. MetaMap analyses biomedical free-text and identifies concepts belonging to the Unified Medical Language System (UMLS), associating each mention with a number of concepts from the UMLS Metathesaurus\footnote{\url{https://www.nlm.nih.gov/research/umls/knowledge_sources/metathesaurus/}} --- which comprises more than 3 million distinct concepts. \par 

Currently, we are working on the relation extraction part. Before trying to develop our own relation extraction algorithm, we tested different open-source natural language processing tools. We started with the tool most related to our needs: SemRep\footnote{\url{https://semrep.nlm.nih.gov/}}. SemRep can detect some relations between UMLS concepts using hand-crafted rules. However, SemRep is based on relations belonging to the UMLS Semantic Network, which relates UMLS Semantic Types --- not UMLS concepts. Therefore, SemRep is set to a lower level of granularity, with respect to the possible relations that can occur between concepts, turning out to be too generic for our task. Besides, SemRep was specifically made for extracting hypernymic propositions. We also tested tools developed for the general-knowledge domain. However, these systems, not having been developed expressly for the biomedical domain, were not satisfying enough for our task. \par 

Being targeted to CDS --- i.e. voted to the extraction of key relations that can facilitate clinical decision making --- our problem setup is fundamentally different from the conventional biomedical setups. Most of state-of-the-art biomedical relation extraction techniques are developed for specific relations, like protein-protein interactions, gene-disease interactions and so on. Therefore, such techniques are targeted to specific areas of the biomedical domain, covering only fractions of it. An example is the i2b2 relation extraction task \cite{uzuner_etal-2011}. In the i2b2 relation extraction task, entity mentions are manually labeled, and each mention refers to one of three classes: `problem', `treatment' and `test'. The extraction task focused on assigning relation types that hold between the three classes mentioned. \par

To resemble real-world CDS tasks, where perfect mentions do not exist and the set of relations must have sufficient coverage of the medical domain, our setup requires the entity mentions to be automatically detected and the set of relations to be as representative as possible of the medical domain. Furthermore, due to the lack of annotated documents available for the task of general medical relation extraction, it is also necessary to collect training data with minimal-to-none labeling effort. \par 

To overcome the lack of labeled data, we are evaluating various distant supervision methods. Distant supervision has become a popular choice for training relation extraction algorithms without using manually labeled data. Regarding the set of relations to be considered, we opted for the relations contained in the UMLS Metathesaurus. Within UMLS, a substantial understanding of the medical domain is included, comprising medical concepts, relations, definitions and so on. Therefore, UMLS relations are the most suited for CDS tasks. \par

To build the relation extractor, we are considering to use bidirectional Recurrent Neural Networks (RNNs). The idea behind bidirectional RNNs is that the output at time \emph{t} may not only depend on the previous elements in the sequence, but also on future elements. Intuitively, this type of RNNs fits naturally to natural language tasks like relation extraction. In the biomedical domain, the use of RNNs for relation extraction tasks is still in its early stages. Therefore, there is large room for improvement and a good margin of novelty that can be introduced --- especially considering the problem setup defined. \par 

\begin{figure}[h]
  \includegraphics[width=70mm]{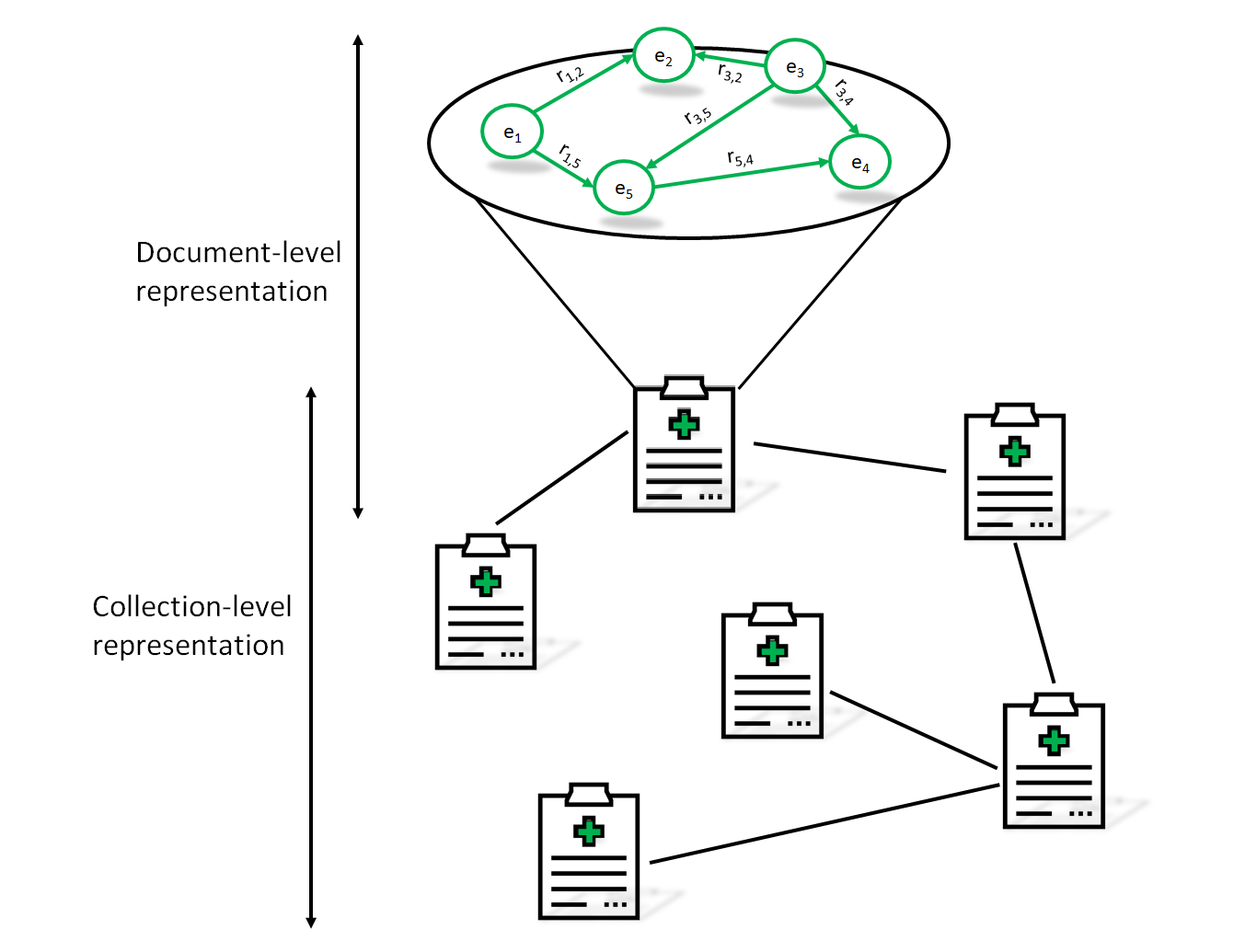}
  \caption{Two-level data representation.}
  \label{fig:collection_network}
\end{figure}

When the creation of documents' semantic networks is performed, the document collection can be considered as a connected graph, where documents are connected by means of semantic relationships determined through the similarity analysis of their semantic networks. Fig~\ref{fig:collection_network} shows this two-level data representation.

\subsection{Case-Based Retrieval}
\label{subsec:case_based_retrieval}
\begin{figure}[h]
  \includegraphics[width=70mm]{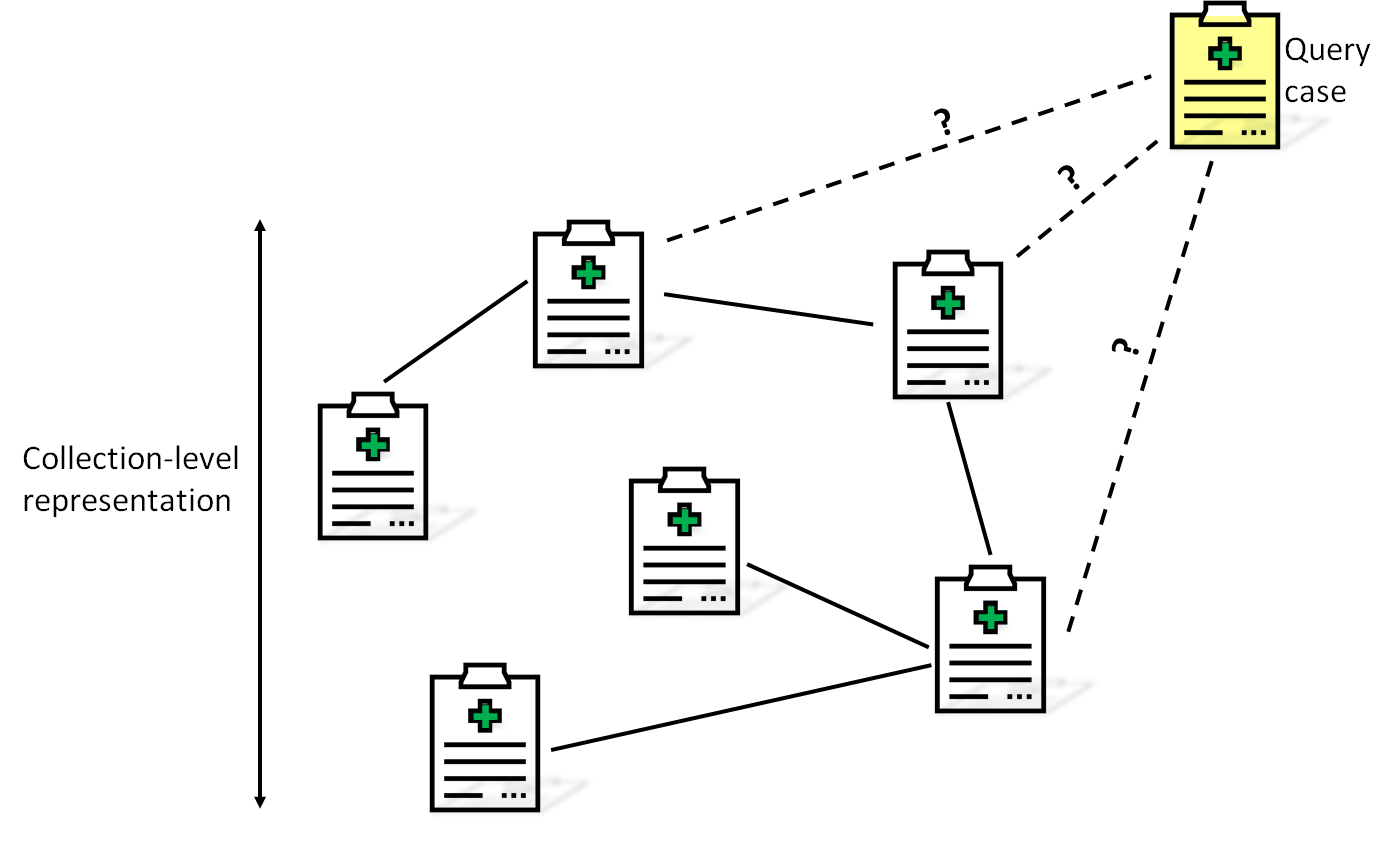}
  \caption{Case-based retrieval as a node insertion task.}
  \label{fig:query_case}
\end{figure}

The second part of the research focuses on leveraging the documents' semantic networks to perform the retrieval of medical literature, given a query case. Due to the nature of the query, graph-based models suit well to be used for case-based retrieval. In fact, the retrieval task can be seen as the insertion of a new node, the query case, in the graph representation of the document collection by finding its closest medical cases, in terms of semantic similarity. Fig~\ref{fig:query_case} shows the concept.  

Since case-based retrieval strongly favors precision over recall, and the most semantically similar nodes to the query case are its nearest neighbors in the graph, the proposed retrieval model is highly suited to address this need. \par

Thus, both graph-based and latent-based models will be investigated for the analysis of documents' semantic networks. We will evaluate which techniques, among the two categories, prove to be the most appropriate for the documents' semantic networks, hence returning the most related medical literature for the given query case. Furthermore, approaches that combine graph-based models with latent-based models will be investigated too, in order to test if considering explicit and implicit information together can enhance semantic similarity techniques and thus, the overall retrieval effectiveness. \par 

\subsection{Evaluation}
\label{subsec:evaluation}
In 2014, NIST's TREC has introduced a CDS search track\footnote{\url{http://www.trec-cds.org/}}. The 2017 track focused on an important use case in clinical decision support: providing useful precision medicine-related information to clinicians treating cancer patients. Therefore, to evaluate the methodology described, we are considering to use the NIST's TREC CDS search track datasets.

\section{Specific research issues}
\label{sec:research_issues}
In this section we highlight some of the possible issues that could affect the research results and push us to consider alternative approaches to those presented above. 

\begin{itemize}
\item The quality of the documents' semantic networks depends strongly on the quality of the IE systems. Therefore, the overall effectiveness of the retrieval system is bounded to these semantic representations of documents.  A minimum level of quality for the networks has to be investigated in order to find the minimum threshold to have competitive results.
\item While representing documents through semantic networks can have the advantage of removing the high amount of noise contained within free-text, it might also have the downside of removing useful information too. Information that could have helped better retrieving relevant documents.
\item In terms of efficiency, the construction of documents' semantic networks can be time consuming. Depending on the size of documents within the collection, the process of creating and then analyzing semantic networks for each document might become prohibitive.
\item If the structure of relational data appears to be not significantly discriminative for computing the similarity between documents' semantic networks, then considering it in addition to semantics might result to be more than necessary.  
\end{itemize}

\bibliographystyle{ACM-Reference-Format}
\bibliography{Marchesin} 

\end{document}